\documentclass[12pt,preprint]{aastex} 
\shorttitle{ Intermediate Asymptotics for Winds and Jets} 
\shortauthors{Heyvaerts and Norman}

\begin{document} 
 
\title{  KINETIC ENERGY FLUX {\it VS} POYNTING FLUX IN MHD WINDS AND
JETS: THE INTERMEDIATE REGIME}
 
\author{ Jean Heyvaerts } 
\affil{Universit\'e Louis Pasteur, Observatoire de Strasbourg 
\altaffilmark{1,4} } 
\email{heyvaert@astro.u-strasbg.fr} 
  
\and 
  
\author{Colin Norman } 
\affil{Department of Physics and Astronomy, Johns Hopkins University \\ 
and Space Telescope Science Institute\altaffilmark{2,3}  } 
\email{norman@stsci.edu} 
\bigskip 
 

\altaffiltext{1}{Observatoire, Universit\'e Louis Pasteur, 
11 Rue de l'Universit\'e, 67000 Strasbourg, France            } 
   
\altaffiltext{2}{Department of Physics and Astronomy, The Johns Hopkins University, Homewood Campus, 3400 North Charles Street, Baltimore, MD 21218} 
   
\altaffiltext{3}{Space Telescope Science Institute, 
3700 San Martin Drive, Baltimore, MD 21218} 
  
\altaffiltext{4}{Visiting Scientist at Space Telescope Science Institute 
and Department of Physics and Astronomy, Johns Hopkins University}

\begin{abstract} 
 
We show that the formal asymptotic limit for all rotating polytropic
axisymmetric perfect MHD flows is a kinetic energy dominated wind
which collimates to paraboloids around the symmetry axis.  We reach
this result by showing that another, {\it a priori} possible, solution with
finite Poynting flux can be excluded on the following physical
grounds: (1) the wind velocity does not exceed the fast mode speed
everywhere and (2) the circumpolar current increases with distance
from the source.

We show that asymptotic hoop stress collimation is mathematically
robust and we give strong arguments why recent 'anti-collimation' claims
are not correct.

However, in practice, due to the very slow logarithmic decline of the
circumpolar current with increasing distance from the source, there is
a broad intermediate regime with significant Poynting flux.  This
intermediate regime, rather than the mathematically exact asymptotic
regime, may well apply along the finite length of the jet. We briefly
discuss peculiarities that would be associated with Poynting jets in
the intermediate regime. Force-free initial conditions in the near
field are most likely to produce such jets, in which most of the
energy flux is electromagnetic.

\end{abstract} 
 
\section{Introduction} 
 
We have previously established (\citet{HN89}) that any stationary 
polytropic axisymmetric magnetized wind will collimate to paraboloids or 
cylinders along the symmetry axis at large distances from the source 
according to whether the electric current brought to infinity per 
hemisphere by the wind is finite or vanishes.  In two companion papers 
\citep{HN2001a, HN2001b} we have presented explicit 
asymptotically-matched solutions for both classical and relativistic 
winds.  Our results however have left open the question of
whether the asymptotic circumpolar current vanishes or not.
We now address this issue which we already 
touched in earlier reports on our work
(\citet{chamonix}, \citet{texas98}). 
 
Previous work has described the transfield equation (\citet{okamoto75} 
\citet{HeinemannOlbert}) that expresses the balance of forces in a 
general magnetized rotating flow in the presence of a gravitational 
field. \citet{blandfordpayne} showed how collimated flows from disks 
could be obtained with a similarity solution associated with a 
particular scale free current and field distribution on a Keplerian 
disk. \citet{ContopoulosLovelace94} have found a similarity solution 
incorporating gravity and pressure, extending and generalizing the 
Blandford and Payne approach, as also did 
\citet{EOstriker97}. Lovelace and collaborators have also considered 
relativistic jets, using the force free approximation
(\citet{Lovelace76}, \citet{Lovelaceetal91}, \citet{Lovelaceetal93}).
In \citet{Lovelace2002} they find collimated Poynting flux jets with
surrounding winds.  Numerical solutions of the steady state rotating
axisymmetric MHD system have been presented by \citet{sakurai85} and
\citet{sakurai87} for the split monopole and magnetized disk. A recent
study of the relativistic case is given by \citet{bogovalov01}.
Collimation along the axis is clearly evident in both cases.  This
result has been strengthened by a number of analytical and numerical
studies (\citet{tsinsauty92a}, \citet{tsinsauty92b},
\citet{sautyetal94}). Shu and collaborators, have analysed
protostellar outflows (\citet{Shuetal94a}, \citet{Shuetal94b},
\citet{NajitaShu94},
\citet{Shuetal95}).  \citet{PelletierPudritz} produced special 
solutions that exhibited focussing and recollimation. Numerical 
results (\citet{Ustyugovaetal95}, \citet{Ustyugovaetal00}, 
\citet{ouyedpudritz97a}, \citet{ouyedpudritz97b}, 
\citet{krasnopolsky}) further confirmed this. 
 
\citet{HN89} have used asymptotic analysis to 
understand the general properties of the shape of the field lines and 
the flow structure at large distances from the source as a function of 
the conserved flow quantities. They have shown that the structure of the 
field and flow at infinity is controlled by the total poloidal 
electric current flowing in the wind about the polar axis. 
In the case where this current 
vanishes, the poloidal field (and flow) lines asymptotically approach 
to parabolae which focus to the polar axis in such a way that both 
$r\rightarrow\infty$ and $(z/r)\rightarrow\infty$ where standard 
cylindrical coordinates ($r, \theta, z$) are used with $z$ along the 
symmetry axis.  An expression giving the shape of these lines at 
infinity far from the polar axis, as a function of the flux function 
was derived. When the circum-polar poloidal current is non-zero, the poloidal 
field (and flow) surfaces have been shown (\citet{HN89}, 
\citet{HN96}) to asymptotically approach to cylinders (possibly 
nested in asymptotically conical flux surfaces).  The asymptotic radii 
of these cylindrical flux surfaces are given as a function of magnetic 
flux in the super-Alfv\'enic regions. In a relativistic 
generalization of our 1989 paper, \citet{ChiuehLiBeg} showed our 
results held not only for the non-relativistic case but also for the 
relativistic case. 

In a recent paper \citet{Okamoto03}
insists that the asymptotic circum-polar current must
vanish and that the asymptotic structure of magnetic surfaces must be
of the conical type.  Our earlier results \citep{HN89} by no means
imply that the asymptotic circum-polar current should assume a
non-vanishing value, contrary to what \citet{Okamoto03} claims. This
point has been made clearly in a number of conference reports
(\citet{chamonix}, \citet{texas98}).  On the contrary, our present
results, which extend and make precise previous arguments published in
these reports, just conclude the contrary, i.e., that the
circumpolar current vanishes aymptotically in the mathematical
sense. Thus, there is no disagreement between us on this point. By
contrast, Okamoto's antifocusing statement is inconsistent with an
asymptotically vanishing circumpolar current.
A non focusing magnetic surface must obviously be a surface such that
$(z/r)$ approaches a finite value at infinite distance from
the origin on that surface, or, if such a limit does not exist,
it is a surface along which $(z/r)$ remains bounded from above.
On such a magnetic surface 
$\rho r^2$ must approach a finite value 
(or remain bounded from below) at infinite
distance from the origin. Indeed, an 
elementary calculation shows that when $(z/r)$ approaches a finite
limit (or is bounded) so does $r \vert \vec{\nabla} a \vert$, 
which implies that the polodal current enclosed in this surface 
approaches a finite value (see \citet{HN89} and \citet{HN96}
for details). If the mathematical asymptotic state,
because of a very slow convergence, is to be reached 
only at distances larger than the actual length of the jet,
there is of course no point to
confront the actual wind structure with this asymptotic
solution. While this is the point made in this paper, it
is not what is meant by \citet{Okamoto03} when he claims conical
asymptotics.

The aim of this paper is precisely to determine whether, in general, the 
asymptotic circumpolar current vanishes or is finite. 
This question cannot be dealt with in the 
framework of a self-similar model, nor of any model which imposes 
a priori constraints on the solution.  In this paper we discuss this 
issue in terms of the properties of the first integrals of the motion, 
which we take as given.  These first integrals are determined 
by boundary conditions and by the  
criticality conditions. We consider both classical and relativistic winds. 
 
The paper is structured as follows. In Section 
\ref{secgeneralproperties} we review the basics of both relativistic 
and non-relativistic stationnary axisymmetric rotating MHD winds.  In
section \ref{secearlier} we summarize results on the asymptotic
structure of MHD winds relevant to the distribution of flux in the
asymptotic domain.  In section \ref{sectionfill} we show that a
necessary condition for solutions with a non-vanishing circumpolar
asymptotic current to be possible is that the function $\alpha (a)
E(a)/ \Omega (a)$ has a minimum value, $I_{sup}$, away from the polar
axis. We show that this non-vanishing asymptotic circumpolar poloidal
electric current is proportional to this minimum value.  The analysis
extends to relativistic winds, for which a non-vanishing asymptotic
proper current must be the minimum value, $K_{sup}$, of $\alpha (a)
(E(a) - c^2) / \Omega (a)$. We next show in section
\ref{sectfastmode} that classical winds which are super fast mode at
infinity must have a current less than ${2 \over 3 }I_{sup}$. A
similar upper bound is derived for relativistic winds.  
Due to the very slow decline of the current, 
the final asymptotic regime may in practice 
not be reached, allowing for intermediate quasi-asymptotic 
regimes. These are discussed in section \ref{secintermediate} where we 
emphasize systems carrying close to the maximum current.  These winds 
split into a separate conical jet of small opening angle and a 
circum-equatorial wind separated by a region where the energy flux is 
almost all in Poynting form.  We summarize 
our results in section \ref{sectionconclusion}. 
 
\section{General Properties of MHD Winds} 
\label{secgeneralproperties} 
 
\subsection{Notation and Definitions} 
\label{subsecnotations} 
 
We study perfect MHD, axisymmetric, stationary, polytropic jets and 
winds. The physical quantities are denoted by their usual symbols, 
$\rho$, $p$, $\vec{v}$, $\vec{B}$ for the density, pressure, velocity 
and magnetic field respectively. The gravitational potential is 
$\Phi_G$.  Any axisymmetric vector quantity can be split into poloidal 
and toroidal components. The magnetic flux $\Phi_m$ through a circle 
of radius $r$ centered on the axis at altitude $z$ is written as 
$\Phi_m = 2 \pi a(r,z)$.  The poloidal part of the magnetic field can 
then be expressed as 
\begin{equation} 
\vec{B}_P = - { 1 \over r} \ {{\partial a}\over{\partial z}} 
\vec{e}_r + { 1 \over r} \ {{\partial a}\over{\partial r}} 
\vec{e}_z 
\label{defa} 
\end{equation} 
The magnetic surfaces, generated by 
the rotation of field lines about the axis, 
are surfaces of constant  $a(r,z)$. The value of $a$ 
suitably labels them. 
 
\subsection{ First Integrals} 
 
A rotating, axisymmetric, stationary, polytropic perfect MHD wind flow 
admits five first integrals of the motion (see for example 
\citet{HeyvMiniato}) $\alpha$, $\Omega$, $L$, $E$ and $Q$ which are 
conserved on magnetic surfaces, $a(r,z)$. For classical winds, these 
first integrals are defined by: 
\begin{equation} 
\rho \vec{v}_P = \alpha(a) \vec{B}_P , 
\label{defalpha} \end{equation} 
\begin{equation} 
\rho v_{\theta} = \alpha (a) B_{\theta} + \rho r \Omega (a)   
\label{defomega} \end{equation} 
\begin{equation} 
L(a) = r v_{\theta} - {r B_{\theta} \over{ \mu_0 \alpha(a)}}. 
\label{defL} \end{equation} 
\begin{equation} 
E(a) = {1\over2}v_P^2 +{1\over2}v_{\theta}^2 +{\Gamma\over \Gamma - 1} 
\ {p\over\rho} +\Phi_{G} - {{r B_{\theta} \Omega(a)} \over{ \mu_0 \alpha(a) }} 
\label{Bern} \end{equation} 
\begin{equation} 
p = Q(a) \rho^{\Gamma} 
\label{polytrop} \end{equation} 
$\Gamma$ is the polytropic index.  
The Alfven radius $r_A(a)$ and the Alfven density $\rho_A(a)$  
are defined by  
\begin{equation} 
r_A^2(a) = L(a)/\Omega(a)  \qquad \qquad  
\rho_A(a) = \mu_0 \alpha^2(a) 
\label{defalfvenquant} \end{equation} 
The Alfv\'en point on a magnetic surface $a$ is at $r = r_A(a)$. The 
density at this point is $\rho_A$.  
The toroidal variables $v_{\theta}$ and $B_{\theta}$ can be 
expressed in terms of first integrals and of the density $\rho$ 
as: 
\begin{equation} 
v_{\theta} = {L \over r} + {\rho\over r}  
\  {L - r^2 \Omega \over \mu_0 \alpha^2 - \rho}, 
\label{vtheta} \end{equation} 
\begin{equation} 
B_{\theta} = {\mu_0 \alpha  \rho \over r}  
\  {L - r^2 \Omega \over \mu_0 \alpha^2 - 
 \rho} . 
\label{Btheta} \end{equation} 
 
\subsection{Relativistic Winds} 
\label{subsecbasicsrelat} 
 
Relativistic flows are characterized by a local Lorentz factor, 
$\gamma$, defined by: 
\begin{equation} 
\gamma = \left(1 - {{(v_{\theta}^2 + v_P^2)} \over{ c^2}}\right)^{-1/2} 
\label{defgamma} \end{equation} 
The proper rest mass density 
is denoted by $\rho$ and the specific momentum (the momentum 
per unit mass)  is 
\begin{equation} 
\vec{u} = \gamma \vec{v} 
\label{specmomentum} \end{equation} 
The proper gas pressure is 
assumed to be  
related to the proper density by Eq.(\ref{polytrop}), where 
$Q$ is constant following the fluid motion. 
We define the function 
\begin{equation} 
\xi = 1 + {\Gamma \over \Gamma -1} {Q \rho^{\Gamma - 1} \over c^2}. 
\label{defxi} \end{equation} 
which is also equal to $(1+\int dP/\rho c^2)$ 
calculated at constant entropy for a polytropic gas. 
We denote  the electric charge density  by $\rho_e$, 
the electric current density  by $\vec{j}$ and  
the mass of the central object by $M_\ast$. 
The special-relativistic equation of motion can be written 
as: 
\begin{equation} 
\gamma \rho (\vec{v} \cdot \vec{\nabla}) 
(\gamma \xi \vec{v}) = 
-  \vec{\nabla} P + \vec{j} \times \vec{B} + \rho_e \vec{E} 
+ \gamma \rho \vec{\nabla} 
\left(\gamma \xi {{G M_{\ast}}\over{R}}\right) 
\label{releqmotion} \end{equation} 
The relativistic form of the laws of mass conservation, 
isorotation, angular momentum conservation 
and Bernoulli involve surface functions 
$E$, $\alpha$, $L$, $\Omega$ and $Q$, which are in this case 
defined  by the relations: 
\begin{equation} 
\gamma \rho \vec{v}_P = \alpha(a) \vec{B}_P , 
\label{defalpharel} \end{equation} 
\begin{equation} 
\gamma (v_{\theta} - r \Omega(a)) = \alpha(a)  B_{\theta} /\rho , 
\label{defOmegarel} \end{equation} 
\begin{equation} 
\gamma \xi r v_{\theta} -{r B_{\theta}\over \mu_0 \alpha(a)} = L(a) , 
\label{defLrel} \end{equation} 
\begin{equation} 
\gamma \xi ( c^2 -{G M_{\ast}\over R}) 
- { r \Omega (a)  B_{\theta}\over \mu_0 \alpha(a)} 
= E(a) 
\label{defErel} \end{equation} 
Note that $E$ in Eq.(\ref{defErel}), now includes the rest mass 
energy.  The rotation rate of the magnetic field, $\Omega(a)$, which 
appears in equations (\ref{defOmegarel}) and (\ref{defErel}), is 
defined in terms of the electric field by: 
\begin{equation} 
\vec{E} = - \Omega (a) \vec{\nabla} a 
\label{omegaandEfield} \end{equation} 
Toroidal variables may be expressed using Eqs.(\ref{defLrel}) and 
(\ref{defOmegarel}): 
\begin{equation} 
r B_{\theta} = \mu_0 \alpha \rho \ \ 
{{L - \gamma r^2 \xi \Omega} 
\over 
{\mu_0 \alpha^2 \xi - \rho}} 
\label{Bthetasimplerel}  \end{equation} 
\begin{equation} 
\gamma \xi v_{\theta} = {L \over r} 
+ {\rho \over r} \ 
{{L - \gamma r^2 \xi \Omega} 
\over 
{\mu_0 \alpha^2 \xi - \rho}} 
\label{vthetasimplerel}  \end{equation} 
Since  $\gamma$ depends on $v_{\theta}$, 
the elimination of the 
toroidal variables in Eqs.(\ref{Bthetasimplerel}) and  
(\ref{vthetasimplerel}) 
is not yet complete. These 
expressions can be substituted in 
the Bernoulli equation (\ref{defErel}), which can then be solved 
to obtain an expression of $\gamma \xi$ in terms 
of the poloidal variables. This 
eventually gives the toroidal 
variables in terms of poloidal variables alone as: 
\begin{equation} 
r B_{\theta} = - \mu_0 \alpha 
{{ L (c^2 - GM_{\ast}/R) - r^2 \Omega E 
}\over{ 
(c^2 - GM_{\ast}/R) (1 - \mu_0 \alpha^2 \xi/\rho) 
-r^2 \Omega^2  }} 
\label{Bthetarel} \end{equation} 
\begin{equation} 
r v_{\theta} = r^2 \Omega \left( 
1 \ \ \ - {{\mu_0 \alpha^2 \xi}\over{\rho}} 
{{ L (c^2 - GM_{\ast}/R) - r^2 \Omega E 
}\over{ 
 (1 - \mu_0 \alpha^2 \xi/\rho) r^2 \Omega E 
- r^2 \Omega L }}\right) 
\label{vthetarel} \end{equation} 
These expressions have to be regular where their denominator vanishes, 
defining quantities which play the role of Alfv\'en density and 
Alfv\'en radius in the relativistic context.

\section{Asymptotics of MHD Winds} 
\label{secearlier} 
 
\subsection{Semantics} 
\label{semantics} 
 
We define important expressions to be used in this paper.  The {\it 
asymptotic domain} consist of all points which are located on their 
own magnetic surface far away from the corresponding Alfv\'en point. 
A magnetic surface is said to be {\it asymptotically parabolic} if the 
limits following this surface of $r$ and $(z/r)$ are both infinite. A 
magnetic surface is said to be {\it asymptotically conical} if $r$ 
approches infinity and $(z/r)$ approaches a finite limit, $\tan 
\psi_{\infty} (a)$. This does not imply that $(z - r \tan 
\psi_{\infty}) $ approaches a finite limit: conical magnetic surfaces 
may have parabolic branches. An asymptotically cylindrical magnetic 
surface is one on which $r$ approaches a finite value, 
$r_{\infty}(a)$.  The ratio $r/r_A$ then also approaches a finite 
limit.  When $r$ becomes infinite on a magnetic surface, the surface 
is said to flare out. 
 
A {\it neutral} or {\it null} magnetic surface is one on which the 
poloidal field vanishes.  The toroidal field also vanishes, since it 
is generated from the poloidal field by rotation.  A neutral surface 
is strictly speaking not a magnetic surface, since it contains no 
magnetic field lines. Null surfaces are sandwiched between regular 
magnetic surfaces, so that the definition still retains meaning in the 
limit.  The immediate vicinity of neutral magnetic surfaces is of 
special interest since they are exceptional regions where electric 
currents can flow in the asymptotic domain (\citet{HN2001a}, 
\citet{HN2001b}).  We refer to these regions as neutral surface 
boundary layers. The polar boundary layer in the vicinity of the polar 
axis is similarly a region where electric currents can flow in the 
asymptotic domain. We call the region away from boundary layers  
the field region. 
 
The total poloidal current, $J = 2 \pi r B_{\theta}/\mu_0$, is the 
electric current through a circle of axis $z$ passing the point ($r,z$). 
For positive $\Omega$ and $\alpha$, $J$ is negative. We shall  
loosely refer to the quantity 
\begin{equation} 
I = - r B_{\theta}/\mu_0  
\label{defI} 
\end{equation} 
as the current.  The total poloidal electric current enclosed in a 
neutral surface is zero. This means that neutral surfaces separate the 
wind in a number of cells in each of which the total current 
separately closes.  We refer to the space between two neighbouring 
neutral magnetic surfaces as being a cell.  In the asymptotic domain, 
current-carrying regions are restricted to regions of small extent 
(\citet{HN2001a},\citet{HN2001b}). 
 
There is a total current that flows about the polar axis which we 
refer to as the {\it circum-polar current}. This circumpolar current 
varies with the distance to the wind source. 
 
\subsection{Relevant Results} 
\label{subsecclassicalasympt} 
 
For $ r \gg r_A$, the azimuthal velocity vanishes while the current 
approaches the value 
\begin{equation} 
I \approx {{\rho r^2 \Omega(a)}\over{\mu_0 \alpha(a) }} 
\label{rBthetaass} \end{equation} 
\citet{HN89} show that $\rho r^2$ is bounded from above.  
The toroidal component of the velocity, $v_{\theta}$, 
approaches zero as $ r$ approaches infinity on any flaring magnetic 
surface. Thus the flow velocity becomes poloidal while $I$ remains 
bounded.  If $(\rho r^2)$ approaches a finite limit, the asymptotic 
structure of the magnetic surfaces consists of a set of conical 
surfaces, inside of which cylindrical magnetic surfaces are 
nested. Such flows convey Poynting flux to infinity.  On cylindrical 
magnetic surfaces none of these results strictly apply.  Here, as in 
our previous papers, we nevertheless assume that the asymptotic 
cylindrical radius on any magnetic surface is much larger than the 
corresponding Alfv\'en radius.  If $\rho r^2$ does converge to zero at 
large distances, the asymptotic structure of magnetic surfaces 
consists of nested paraboloids \citep{HN89}. No Poynting flux reaches 
infinity and all the energy emerges in kinetic form.   
 
A detailed asymptotic solution in terms of given first-integrals has been 
presented in the companion papers \citep{HN2001a, HN2001b} 
for both classical and relativistic winds. We have shown that electric 
currents are distributed in the asymptotic domain in boundary layers 
about the pole and in the vicinity of neutral magnetic surfaces, with 
very little current density flowing outside these regions.  Detailed 
solutions for the inner structure of these current-carrying boundary layers 
were also discussed. 
Let $a_n$ be the flux variable of some neutral magnetic surface.  If 
there is a mass flux along the surface, the first integral $\alpha$ 
diverges for $a \rightarrow a_n$, as can be seen from 
Eq.(\ref{defalpha}).  We have established that outside these 
boundary layers the transfield equation takes the simple form 
\begin{equation} 
I = I_{\infty} (b) 
\label{Iofb}\end{equation} 
where $b$ labels surfaces orthogonal to magnetic 
surfaces. Specifically, the label $b$ is taken to be the value of 
the $z$-coordinate on this orthogonal trajectory at the polar 
axis. Variables $a$ and $b$ could, in principle, be taken as space 
coordinates replacing $r$ and $z$, with $a$ playing the role of an 
angular variable and $b$ the role of a radial variable. 
 
The current variable, $I$, in general depends on both $a$ and $b$. 
Eq.(\ref{Iofb}) expresses the fact that its dependance on $a$ 
almost completely disappears in the field-regions of the asymptotic domain, outside the 
current carrying boundary layers.  More precisely, $I_{\infty} (b)$ is 
only piecewise constant, reversing sign when crossing from one cell to 
another. This follows from consideration of the equilibrium of the 
neutral sheet boundary layers where the dominant toroidal magnetic 
pressure on each side must balance the other. So, $B_{\theta}$, and 
$I$ reverse sign as neutral magnetic surfaces are crossed.  This 
means that the absolute value of the total poloidal current $\vert 
I_{\infty}(b) \vert$ passes unchanged from one to the next cell. 
 
Winds therefore could come in two separate classes: 
Poynting jets approaching a finite value of $\vert I_{\infty}(b) \vert$ 
or kinetic winds having asymptotically vanishing 
current. Whether both classes actually occur in nature remains to be 
discussed. 
 
Outside the current-carrying boundary layers, the Lorentz force $ 
\vec{j} \times \vec{B}$ vanishes since $\vec{j}_P$  and 
$\vec{B}_P$ become both small.  Inside boundary layers, 
other forces balance the Lorentz force \citep{sautyetal99}.
The gas pressure is the most 
likely additional force in the case of polytropic flows since its 
decline with $r$ is the slowest.  The polar boundary layer has the 
structure of a column pinch, while the neutral surface boundary layers 
have the structure of sheet pinches (\citet{HN2001a}). 
 
\subsection{Distribution of Flux on Orthogonal Trajectories:  
Classical Winds} 
\label{fluxonorthogtrajclass} 
 
For large values of $r/r_A$, it is possible to determine the 
distribution of flux along trajectories orthogonal to magnetic 
surfaces by quadrature. Denoting the terminal velocity on surface $a$ 
by $v_{\infty} (a)$, equations (\ref{defalpha}), (\ref{Bern}) and 
(\ref{Btheta}) simplify to the form: 
\begin{equation} 
\rho r  v_{\infty} (a) =  \alpha \vert \vec{\nabla} a \vert  
\label{defalphaasympt} 
\end{equation} 
\begin{equation} 
\mu_0 I =  + {{\rho r^2 \Omega}\over{\alpha}}  
\label{Iasympt} 
\end{equation} 
\begin{equation} 
E = {{v_{\infty}^2}\over{2}} + {{I \Omega}\over{\alpha}} 
\label{Bernasympt} 
\end{equation} 
Eliminating $v_{\infty}$ and $\rho$ we obtain an expression for $r 
\vert \vec{\nabla} a \vert$ in terms of the first integrals and $I$: 
\begin{equation} 
r \vert \vec{\nabla} a \vert  
= {{ \sqrt{2} \mu_0 \vert I \vert }\over{\Omega}} 
\left(E - {{I \Omega}\over{\alpha}}\right)^{1 \over 2} 
\label{rgradaofIgeneral} 
\end{equation} 
Let the curvilinear abcissa along the trajectories orthogonal to the 
magnetic surfaces be denoted by $\sigma$, conventionally increasing 
from pole to equator. Then $\vert \vec{\nabla} a \vert = da/d\sigma$, 
and the relation (\ref{rgradaofIgeneral}) becomes: 
\begin{equation} 
{{d \sigma}\over{r}} =   
{{\Omega(a) da}\over{\mu_0 I(a,b) \sqrt{2}  
\sqrt{E - I(a,b) \Omega(a)/\alpha(a) }   }} 
\label{fluxversuscurvilinear}        
\end{equation} 
This relation holds true both in the field-regions, where $I(a,b)$ 
does not depend on $a$, and in current carrying boundary layers, where 
this dependence is rather strong. 
 
The actual form of the quadrature relation 
(\ref{fluxversuscurvilinear}) depends on the actual shape of the 
orthogonal trajectories to magnetic surfaces.  These trajectories have 
been shown to be approximately circles when $I(a,b)$ approaches a 
non-vanishing limit $I_{\infty}$ (\citet{HN2001a}, \citet{HN2001b}). 
This result has been extended in a WKB sense to the case when $I$ 
slowly varies with $b$.  Asymptotically cylindrical surfaces are 
approximately orthogonal to circles too.  The position on such a circular 
orthogonal trajectory is specified by a latitude angle $\psi$ from the 
equator, so that 
\begin{equation} 
z = r \  \tan (\psi(a)) 
\label{eqofcones} \end{equation} 
For approximately circular orthogonal trajectories 
Eq.(\ref{fluxversuscurvilinear})  
integrates to 
\begin{equation} 
\tan (\psi (a)) = \sinh \left( 
\int_a^{a_1} {1 \over \sqrt{2} \mu_0 I(a,b)} 
{\Omega(a') da' \over 
\sqrt{ E(a') - I(a,b)  \Omega(a') / \alpha(a') }}\right) 
+ \tan (\psi (a_1)) 
\label{soltgpsiofa}  \end{equation} 
where $a_1$ is some reference flux.  Along the orthogonal trajectory 
$b$ the current $I(a,b)$ vanishes near the polar axis proportionally 
to $a$ and causes $\tan(\psi(a,b))$ to diverge as $a$ approaches 
zero. Outside boundary layers, $I(a,b)$ becomes a 
function $I_{\infty}(b)$, essentially independent of $a$ at constant $b$, and the 
solution (\ref{soltgpsiofa}) simplifies. It will be 
shown below that the integral in Eq.(\ref{soltgpsiofa}) converges at a 
neutral surface. If, moreover, we neglect the flux in neutral boundary 
layers the integration in Eq.(\ref{soltgpsiofa}) can be extended from 
the equator, at flux $a = A$, where $\tan \psi(A) = 0$, to the surface 
$a$ with a  constant value of $\vert I(a,b) \vert = 
I_{\infty}$.  This gives an approximate expression for $\psi(a)$: 
\begin{equation} 
\tan (\psi (a)) = \sinh \left( 
\int_a^{A} {{1 }\over{ \sqrt{2} \mu_0 I_{\infty}(b) }} 
\ {{\Omega(a') da' }\over{  
\sqrt{ E(a') - I_{\infty}(b) \Omega(a') / \alpha(a')} }}\right) 
\label{soltgpsiofawithoutneutrallayers} 
\end{equation}

\subsection{Distribution of Flux on Orthogonal Trajectories: 
Relativistic Winds.} 
\label{subsecdistribfluxrelativistic} 
 
A similar analysis can be done in the case of relativistic winds 
(\citet{HN2001b}). In the field-regions the quantity $K$, defined by: 
\begin{equation} 
K = {{ r}\over{\mu_0 c}} 
\sqrt{ c^2 B_{\theta}^2 - \Omega^2 \vert \vec{\nabla} a \vert^2 } 
\label{defKrelat} 
\end{equation} 
becomes a constant. We refer to $K$ as the proper total current 
\citep{HN2001b}. $K$ is $r$ times the asymptotic value of the 
electromagnetic invariant $(B^2 - E^2/c^2)^{1/2}$.  It reduces in the 
classical limit to the poloidal current $I$ (see Eq.(\ref{IandK})). In 
field-regions $K$ becomes independent of $a$ along any orthogonal 
trajectory, giving: 
\begin{equation} 
K = K(b) 
\label{constancyK} 
\end{equation} 
The asymptotic relativistic  
Bernoulli equation can be written as: 
\begin{equation} 
u_{P\infty} \equiv \gamma_{\infty} v_{P \infty} = c  
{{\sqrt{E^2 -(c^2 +{K \Omega / \alpha})^2 } 
}\over{ 
(c^2 +{K \Omega / \alpha})}} 
\label{Bernassrel} \end{equation} 
Note that in relativistic dynamics 
$E$ contains the rest-mass specific energy, 
$c^2$, so that $E_{class}= E_{relat} - c^2$.  
The relativistic form of Eq.(\ref{fluxversuscurvilinear}) is: 
\begin{equation} 
{{d \sigma}\over{r}} =   {{\Omega(a)  
\sqrt{c^2 - v_{\infty}^2(a)} \ da}\over{\mu_0 c K(a,b)  
\ v_{\infty}(a) }} 
\label{fluxversuscurvilinearrelat} 
\end{equation} 
For relativistic winds, as for classical ones, the orthogonal 
trajectories to magnetic surfaces at large distances 
closely resemble circles, in which case  
Eq.(\ref{fluxversuscurvilinearrelat}) can be integrated 
to find  the latitude $\psi(a)$ of the magnetic surface $a$ 
on the orthogonal trajectory $b$. This gives: 
\begin{equation} 
\tan \left( \psi (a,b) \right) = \sinh \left( 
\int_a^{a_1} {1 \over \mu_0 c \ K(a,b)} 
{{ \Omega(a')  da' (c^2 + K(a,b) \Omega(a')/\alpha(a')) 
}\over{ 
\sqrt{E^2(a') - \left( c^2 + K(a,b) \Omega(a') /\alpha(a')\right)^2 } 
  }} 
\right) + \tan \psi(a_1 , b) 
\label{soltgpsiofarel} \end{equation} 
where $a_1$ is a reference flux in the same cell as $a$ itself. In the 
case of dipolar symmetry, only one cell would be present per 
hemisphere. Neglecting the flux in the equatorial boundary layer, 
$a_1$ is the equatorial flux, $A$, and $\tan \psi(A,b)$ 
vanishes. Outside boundary layers, $K(a,b)$ becomes a 
constant $K_{\infty}(b)$, independent of $a$ and the solution given by 
Eq.(\ref{soltgpsiofarel}) simplifies as: 
\begin{equation} 
\tan \left( \psi (a,b) \right) = \sinh \left( 
\int_a^{A} {1 \over \mu_0 c \ K_{\infty}(b) } 
{{ \Omega(a')  da' (c^2 + K_{\infty}(b) \Omega(a') /\alpha(a')) 
}\over{ 
\sqrt{E^2(a') - \left( c^2 + K_{\infty}(b) \Omega(a') /\alpha(a')\right)^2 } 
  }} \right)  
\label{soltgpsiofarelatwithoutneutrallayers} 
\end{equation}

\section{The Circumpolar Current} 
\label{sectionfill} 
 
\subsection{Mixed Cylindrical-Conical Asymptotics} 
\label{consistencyconescylinders} 
 
When $I(a,b)$ converges to zero the solution, which consists of nested 
paraboloids, fills all space.  We have shown in \citet{HN2001a} that, 
conversely, if $\vert I(a,b)\vert$ approaches a non-vanishing value at 
infinity, the asymptotic structure of magnetic surfaces consists of 
cylindrical surfaces, possibly nested in conical ones.  These two 
geometries seem to clash because their common existence implies the 
presence of a big spatial region with no magnetic surface.  between 
the last cylindrical surface and the first conical surfaces. A smooth 
connection is nevertheless possible.  Let $a_*$ be the flux  
at which the magnetic surfaces switch from being cylindrical to being 
conical.  The solution is continuous if $r_{\infty}(a)$ approaches an 
infinite limit as $ a \rightarrow a_*$ and if $\psi(a)$ approaches 
$\pi/2$, as $a$ approaches $a_*$ from above.  Thus, the difficulty of 
matching cylindrical to conical asymptotics is not sufficient 
to eliminate the idea that the total circumpolar current 
could remain finite.  The relevant question is rather whether the 
wind's physics allows this. 
In a cylindrical region, $r_{\infty}(a)$ cannot approach infinity 
uniformly as $a$ approaches $a_*$.  Indeed, the limiting magnetic 
surface has the character of a paraboloidal surface described by $z = 
F_*(r)$ . Any asymptotically cylindrical magnetic surface with 
terminal radius $r_{\infty}(a)$ is nested inside $ 
z=F_*(r_{\infty}(a))$. Since $F_*(r)$ diverges with $r$ , the limit 
$r_{\infty}(a)$ can obviously not be reached uniformly. 
For future reference let us note that the terminal radius, 
$r_{\infty}(a)$ is given by: 
\begin{equation} 
r_{\infty}(a) = r_1 \exp \left(  
\int^a_{a_1} {1 \over \sqrt{2} \mu_0 I_{\infty}} 
{\Omega(a') da' \over 
\sqrt{ E(a') - I_{\infty}  \Omega(a') / \alpha(a') }} \right) 
\label{rinftyofa} 
\end{equation} 
 
\subsection{Asymptotic Current for Classical Winds}  
\label{subsecalphaEoverOmegaclass} 
 
Eqs.(\ref{soltgpsiofa}) and (\ref{soltgpsiofarel}) give the flux 
distribution explicitly in the field regions.  Separate solutions have 
been obtained in current-carrying boundary layers and matched to the 
field-region solutions. For classical winds, a smooth asymptotic 
matching implies a relation, which we refered to as the Bennet 
relation, between the current $I(b)$ and the axial density at the 
distance $b$, $\rho_0(b)$: 
\begin{equation} 
{{\Gamma}\over{\Gamma -1}} \ Q_0 \rho_0^{\Gamma -1} (b) = 
{{I(b) \  \Omega_0}\over{\alpha_0}} 
\label{bennetclass} 
\end{equation} 
Matching also imposes the  relation:  
\begin{equation} 
{{\lambda (n_0(b))}\over{n_0^{\Gamma-1}(b) }} = 
(2 - \Gamma) \ \ln (n_0(b))  \ + 
\ \ln \left( {{4 b^2}\over{\ell^2}} \right) 
\label{matchingrelation}  
\end{equation} 
Here $n_0(b) = \rho_0(b) / \rho_{A0} $ is a dimensionless  
measure of the 
axial density and  
$\rho_{A0} $ is the Alfv\'en density on the polar field line. 
The reference length $\ell$ is  defined by 
\begin{equation} 
\ell^2 = {{\Gamma}\over{\Gamma -1}}  
{{Q_0 \rho_{A0}^{\Gamma -1} }\over{  \Omega_0^2}} 
\label{defell} 
\end{equation} 
and $\lambda$ is the following integral, which
depends on $I(b)$, that is, by Eq.(\ref{bennetclass}), on $n_0(b)$ 
or $\rho_0(b)$
\begin{equation} 
\lambda = \int_0^A 
{{ \Omega(a') \  \sqrt{E_0} }\over{ 
\Omega_0 \ \sqrt{ E(a')- I(b) \Omega(a') / \alpha(a')} }} \ {{da'} 
\over{a_0}} 
\label{factorlambda} \end{equation} 
As $b$  approaches infinity, some term of Eq.(\ref{matchingrelation}) 
has to balance the divergence of the logarithmic term on the 
right. If $I(b)$ asymptotically vanishes, so does 
$\rho_0(b)($Eq.(\ref{bennetclass})), and $\lambda$ approaches a 
finite value. In this case, Eq.(\ref{matchingrelation}) is satisfied 
with $\rho_0(b)$ approaching zero, consistent with 
Eq.(\ref{bennetclass}). 
If, by contrast, $I(b)$ approaches a finite value, $I_{\infty}$, so 
does $n_0(b)$ (Eq.(\ref{bennetclass})).  The divergence of the second 
logarithmic term on the right hand side of  
Eq.(\ref{matchingrelation}) can then 
only be matched by a divergence of $\lambda(n_0(b))$. As discussed in 
\citet{HN2001a}, this implies that, as $b$ approaches infinity, the 
circumpolar poloidal current $I(b)$ rises to its maximum allowed 
value. This value is the absolute minimum of the function $(\alpha 
E/\Omega)$, reached away from the polar axis at $a*$
(see Fig.\ref{fig1}). Then: 
\begin{equation} 
\vert I_{\infty} \vert = I_{sup} =  
{\mathrm{Min}}_{a \ne 0} \  
\vert {{\alpha (a) E(a)}\over{\Omega (a)}} \vert 
\label{alphaEoverOmegaclass} 
\end{equation} 
For this current both  
$\tan(\psi_{\infty}(a))$  
and the terminal radius $r_{\infty}(a)$  
would simultaneously diverge as $a*$ is approached. 
 
From Eq.(\ref{rinftyofa}) we see that $\vert I \vert$ should  
in this case increase 
with $b$. However, this result gives a direct contradiction since the 
circumpolar current must close back to the source {\it via} a low 
current density flow in the field-region. This picture implies that 
$\vert I(b) \vert$ must approach its limit from above. A decreasing 
$\vert I(b) \vert$ accelerates the outflow by the coiled spring force 
associated with the gradient of the toroidal magnetic field pressure. 
A monotonically decreasing $\vert I(b) \vert$ corresponds to 
continued, though weakening, acceleration from the wind source to 
infinity.  An increasing $\vert I(b) \vert$ is associated with a wind 
flow undergoing continuous deceleration and thus the Poynting flux 
increases with distance. Therefore we conclude on these  
physical grounds that the circumpolar current cannot remain finite
at infinity. 
 
\subsection{ Asymptotic  
Proper Current for Relativistic Winds.} 
\label{subsecalphaEoverOmegarelat} 
 
For relativistic winds, an entirely similar analysis  
\citep{HN2001b} 
has shown that relations similar to Eq.(\ref{bennetclass}) and 
(\ref{matchingrelation}) hold true, namely 
\begin{equation} 
{{\Gamma }\over{\Gamma - 1}} Q_0 \rho_0^{\Gamma -1}(b) =  
{{\Omega_0 K(b)  }\over{ 
\alpha_0 }} 
\label{bennetrelat}  
\end{equation} 
\begin{equation} 
{{\lambda_{r} (n_0(b))}\over{n_0^{\Gamma-1}(b) }} = 
(2 - \Gamma) \ \ln (n_0(b))  \ + 
\ \ln \left( {{4 b^2}\over{\ell^2}} \right) 
\label{matchingrelationrelat}  
\end{equation} 
where $\lambda_{r}$, defined by 
\begin{equation} 
\lambda_{r} = \int_0^A \ {{da'}\over{a_0}} 
{{ \Omega(a') \sqrt{c^2 + K(b) \Omega(a') /\alpha(a')}  
 }\over{ 
 \sqrt{ E^2(a') - (c^2 + K(b) \Omega(a') /\alpha(a'))^2 } 
}} \  
{{ \sqrt{ E_0^2 - (c^2 + K(b) \Omega_0/\alpha_0)^2 } 
}\over{ 
\Omega_0 \sqrt{c^2 + K(b) \Omega_0/\alpha_0 } 
}} 
\label{factorlambdarelat}  
\end{equation} 
is a function of $n_0(b)$, since $K(b)$ 
is given by Eq.(\ref{bennetrelat}). 
Similar arguments then show that, if $K$ is to approach a finite limit, 
its value should be the absolute minimum of the function 
$(\alpha (E - c^2)/\Omega)$, reached at a non-zero 
regime change flux, $a_*$. 
So, if $K_{\infty}$ is not zero: 
\begin{equation} 
\vert K_{\infty} \vert = K_{sup} =  {\mathrm{Min}}_{a \ne 0} \  
\vert {{\alpha(a) (E(a) - c^2)}\over{\Omega(a)}} \vert 
\label{alphaEoverOmegarelat} 
\end{equation} 
This limit, however, can only be approached from below, again leading  
to a physical contradiction. Although $K$ is not directly related to 
the poloidal current $I$ anymore, it 
is possible to see that, if the wind source is the only 
current source, $K$ should be a decreasing function of $b$. 
Indeed, the asymptotic limit of $K$, $K_{\infty}(b)$,  
is related to the asymptotic 
current $I_{\infty}(a,b)$ enclosed in surface $a$  
at distance $b$  by  \citep{HN2001b} 
\begin{equation} 
I_{\infty}(a,b) = \gamma_{\infty}(a,b) K_{\infty}(b) 
\label{IandK} 
\end{equation} 
where $\gamma_{\infty}(a,b)$ is the Lorentz factor on surface 
$a$ at distance $b$. Again, $I_{\infty}(a,b)$  
decreasing with $b$, at constant $a$, 
implies plasma acceleration, since the field-aligned  
poloidal Lorentz force $\vec{j}_P \times \vec{B}_{\theta}$ 
is then in the sense of the motion. This causes  
$\gamma_{\infty}(a,b)$ to increase \citep{begli94}.
Thus, $I_{\infty}(a,b)$ decreasing with $b$ implies  that 
$K_{\infty}(b)$ is also decreasing with $b$. 
Solutions with a finite asymptotic value of $\vert K \vert$, 
which can be approached only by increasing values of $\vert K \vert$, are  
therefore physically inconsistent. The circumpolar proper current 
cannot be finite at infinity for relativistic winds.

\subsection{Asymptotic Current:  
Progressive Deconfinement} 
\label{subsecdeconfinement} 
 
That the circumpolar current can only be zero or given by 
Eq.(\ref{alphaEoverOmegaclass}) can also be understood by the 
following analysis. Let us suppose that an external pressure 
independent of $z$ confines the jet in such a way that it has 
cylindrical geometry. At low confining pressure, its structure, which 
is entirely determined by the first integrals and by the boundary 
condition, will consist of a circumpolar current-carrying channel 
surrounded by a large, almost pressureless, field region carrying very 
little current density. In this latter region, the current enclosed in 
a magnetic surface $a$ is almost independent of $a$ and equal to 
$I$. This field-region value of $I$ depends on the external confining 
pressure $P_{ext}$.  We calculate explicitly in 
Appendix (\ref{appdeconfinecylindrical}) the relation between $I$ 
and $P_{ext}$, both when 
the edge of the jet is either a magnetic surface or a null 
surface. The conclusion of this analysis is that $I$ approaches either 
zero or $I_{sup}$ ($K_{sup}$ in the relativistic case) as $P_{ext}$ 
decreases to zero. 
 
\subsection{Asymptotic Current: Conclusion} 
\label{concudeIequalzero} 
 
The discussions in subsections (\ref{subsecalphaEoverOmegaclass}) and 
(\ref{subsecalphaEoverOmegarelat}) have clearly shown that if the wind 
extends over very large distances, the amount of circumpolar electric 
current should approach zero as $z$ approaches infinity.  
When this terminal regime is reached, 
all of the wind energy is in kinetic form.  As shown in \citet{HN89}, 
\citet{HN2001a} and \citet{HN2001b}, the magnetic surfaces are then 
paraboloidal. 

\section{Characteristic MHD Speeds}
\label{sectfastmode}

\subsection{ Terminal Speed {\it vs} Fast Mode Speed}
\label{subsecsuperfast}

In the absence of external pressure, a wind should eventually
achieve, on each field line, a terminal velocity exceeding all the MHD
characteristic velocities.  We now show that the fully asymptotic wind
regime cannot satisfy this condition if the wind is to carry a finite
circumpolar current. Specifically, super fast mode velocities cannot
be reached in the vicinity of the regime-change surface $a_*$ in the
field-region. As discussed above, the terminal velocity has to vanish
at this surface and, for a classical wind, the circumpolar current
should equal $I_{sup}$, the minimum value of the function $(\alpha
E/\Omega)$.  Does the terminal velocity, though vanishing, remains
larger than the fast mode speed?  A necessary, though not sufficient,
condition for the flow remaining super-fast-mode is that the fast mode
velocity itself approaches zero as $a$ approaches $a_*$.

The position and velocity at the fast critical point are obtained
from the Bernoulli equation (\ref{Bern}). Let us write it,
for a given magnetic surface $a$,
in the form ${\cal B}(r, \rho) = 0$. The critical points and associated
velocities are obtained by solving the system
\begin{equation}
r {{\partial {\cal B}}\over{\partial r}} =
-{{1}\over{2}}\,{{\alpha^2}\over{\rho^2}}\,\,r \,\,{{d
B_P^2}\over{dr}} - r {{d\Phi_G}\over{dr}} + \Omega^2  r^2 +
{{\rho_A^2}\over{r^2}} {{L^2 - \Omega^2 r^4 }\over{(\rho_A -
\rho)^2}}= 0
\label{critr} \end{equation}
\begin{equation}
- \rho {{\partial {\cal B}}\over{\partial \rho}} = - {{\alpha^2
B_P^2}\over{\rho^2}} + \Gamma Q\rho^{\Gamma - 1} + {{\rho
\rho_A^2}\over{r^2}}\, {{(L - \Omega r^2)^2}\over{(\rho_A - \rho)^3}} = 0
\label{critrho} \end{equation}
The velocity at the fast point is then, for classical winds :
\begin{equation}
v_f^2 = \Gamma Q \rho^{\Gamma -1}
+ \Omega^2
{{\rho \rho_A^2 (r^2 -r_A^2)^2}\over{r^2 (\rho_A - \rho)^3 }}
\label{vfastclassgeneral} \end{equation}
A lower bound to this is
\begin{equation}
v_f^2 \ge \Omega^2 {{\rho \rho_A^2 (r^2 -r_A^2)^2}\over{r^2 (\rho_A - \rho)^3 }}
\label{limittovf}
\end{equation}
which, from Eqs. (\ref{Btheta}) and (\ref{defalpha})
means that $v_f^2 \ge (B_P^2 + B_{\theta}^2)/\mu_0 \rho$.
A weaker, but accurate, lower bound for $v_f^2$ is $B_{\theta}^2/\mu_0 \rho$,
which is obtained in the limit $r \gg r_A$ and $\rho \ll \rho_A$.
Using Eq.(\ref{rBthetaass}), the inequality (\ref{limittovf})
reduces, in this limit, to
\begin{equation}
v_f^2 \ge \vert {{I \Omega}\over{\alpha}} \vert
\label{limitvfbyI}
\end{equation}
The fluid terminal velocity $v_{\infty}(a)$, given by
Eq.(\ref{Bernasympt}), cannot remain larger than
${I \Omega}/{\alpha}$ if $v_{\infty}$ is to approach zero
as $a$ approaches $a_*$, since
$\vert I \vert $ is supposedly close to a non-zero value  and
$\alpha$ supposedly remains finite. Note however that this
would not be so if $a_*$ were a neutral magnetic surface, where
$I$ vanishes and $\alpha$ diverges.
A more general argument is as follows. Eq.(\ref{Bernasympt}) and
the inequality (\ref{limitvfbyI}) indicate that
the wind can only be super fast mode if :
\begin{equation}
\vert I \vert \le {2 \over 3} \ {{\alpha E}\over{\Omega}}
\label{limitIsuperfast}
\end{equation}
In the field-regions of the asymptotic domain, $\vert I \vert$ is a
constant, $I_{\infty}$. $\vert I \vert$ should be less than $I_{sup}$,
the minimum value of $\vert \alpha E /\Omega\vert$, otherwise, from
Eq.(\ref{Bernasympt}), $v_{\infty}^2$ would not be positive
everywhere.  If it is required that the wind be everywhere super-fast
mode at large distances, inequality (\ref{limitIsuperfast}) shows that
the upper limit $I_{sup}$ cannot be reached.  In particular, the
cylindrical asymptotics regime,for which  $I_{\infty} = I_{sup}$, is not
consistent with the wind being everywhere super-fast-mode.  Winds in
an intermediate regime, which deliver a circumpolar current very close
to the maximum $I_{sup}$, should still be in the sub-fast mode regime.
Unless the environment of the jet makes it possible that its velocity
does not exceed the fast mode speed asymptotically, the electric
current should eventually fall below ${2 \over 3} I_{sup}$ at very
large distances. This argument again points to the conclusion that the
completely asymptotic regime should have a vanishingly small
circumpolar electric current.

\subsection{Fast Point in the Relativistic Regime}

Similar conclusions are found in the case of relativistic winds.  The
location and velocity of the fast point can be found as in the
classical case.  Neglecting gravity and the gas entropy, the
relativistic Bernoulli equation takes the form
\begin{equation}
\left( c^2 + {{\rho r^2 \Omega^2}\over{\mu_0 \alpha^2}} \right)^2
\left( 1 + {{u_p^2}\over{c^2}} \right) = E^2
\label{Bernassrelatgnul}
\end{equation}
Assuming the shape of field lines is given, Eq.(\ref{defalpharel})
turns Eq.(\ref{Bernassrelatgnul}) into an equation for $u_p$,
${\cal{B}}(u_p, r) = 0$.  At critical points the
differential of ${\cal{B}}$ vanishes. The position, $r_f$, of the fast
point is then given by :
\begin{equation}
{{\partial (B_p r^2)}\over{\partial s}} = 0
\label{posfastpointrelat}
\end{equation}
where $s$ is the curvilinear abcissa along a field line.
Eq.(\ref{posfastpointrelat}) can be satisfied at large distances in
any particular geometry compatible with the Bernoulli equation
(appendix (\ref{appendcritatinftyclass})). For an initial monopolar
field at the source of a highly magnetized, relativistic wind the fast
point is at a large distance from the light cylinder
\citep{beskin98}. The specific momentum at the fast point is obtained
from the condition that $\partial {\cal{B}}/\partial u_p$
vanishes. The Lorentz factor associated with the poloidal velocity at
the fast point is:
\begin{equation}
\gamma_{pf} =  \left({{E}\over{c^2}} \right)^{{1 \over 3}}
\end{equation}
Eq.(\ref{Bernassrelatgnul}) can be solved when the flow is in the
asymptotic domain, since $\rho r^2$ becomes a constant at large
distances, resulting in Eq.(\ref{Bernassrel}) which relates the
asymptotic specific momentum to the proper total current K.
A condition for the flow to be super-fast-mode, similar to the
inequality (\ref{limitIsuperfast}), can be derived for
relativistic flows.  The fast mode speed, $c_f$, for a diffuse highly
magnetized medium where the Alfv\'en speed $c_A = (B^2/\mu_0
\rho)^{1/2})$ may exceed the speed of light, is given by
\begin{equation}
{1 \over c_f^2} = {1 \over c^2} + {1 \over c_A^2}
\label{fastspeedrelat}
\end{equation}
The Alfv\'en speed associated with the total field, $c_A$, is larger
than the the Alfv\'en speed associated with the toroidal field, $c_{A
\theta}$.  This gives for $c_f$ the following inequality:
\begin{equation}
c_f^2 \ge {{c_{A \theta}^2 c^2}\over{ c^2 + c_{A \theta}^2}}
\label{lowerboundrelatcf}
\end{equation}
The proper current $K$ is related to $I$ and $\gamma$
by Eq.(\ref{IandK}) and can be expressed in terms of $\rho$ by
\citep{HN2001b}:
\begin{equation}
K = {{\rho r^2 \Omega}\over{\mu_0 \alpha}}
\label{Kandrho}
\end{equation}
Using these relations, the inequality (\ref{lowerboundrelatcf}) can be
transformed into the following inequality for the specific momentum
$u_f= c_f \gamma(c_f) $ associated with $c_f$:
\begin{equation}
u_f^2 \ge E^2 {{ K \Omega / \alpha}\over{
\left( c^2 + K \Omega / \alpha\right)^2 }}
\label{inequalforuf}
\end{equation}
A necessary condition for the relativistic flow  to be
super-fast-mode is that the terminal specific momentum of
the wind flow, $u_{P \infty}$, given by Eq.(\ref{Bernassrel}),
exceeds the lower bound to  the asymptotic value of
$u_f$, given by Eq.(\ref{inequalforuf}).
This can be written as:
\begin{equation}
c^2 \left( E^2 - \left( c^2 + {{K \Omega}\over{\alpha}} \right)^2 \right)
\ge  E^2 {{K \Omega}\over{\alpha}}
\label{ineqforsuperfastrelat}
\end{equation}
Now, if the asymptotic flow regime is to have a finite value of the
circumpolar proper current $K$, the latter must be the minimum value
of $\alpha (E - c^2)/\Omega$, reached at some 
non-zero $a_*$.  At this
particular value of $a$, the inequality (\ref{ineqforsuperfastrelat})
cannot be satisfied, since its left side vanishes whereas its right
side is strictly positive.  Expanding (\ref{ineqforsuperfastrelat})
for $E$ close to $c^2$ gives again the inequality
(\ref{limitIsuperfast}) for classical dynamics. Therefore,
relativistic winds in an intermediate regime having a circumpolar
current very close to the maximum $I_{sup}$ should still be in the
sub-fast mode regime in some regions.
Since the fast mode critical surface
is unlikely to be found far in
the asymptotic domain, this situation should be exceptional. 
By contrast, the completely asymptotic relativistic regime
should have a vanishingly small circumpolar electric current.

\section{Intermediate Asymptotic Regime} 
\label{secintermediate} 
 
\subsection{Is the Mathematical Asymptotic Regime ever Reached?} 
\label{subsecasymptquestioned} 
 
In the fully asymptotic regime, the currents $I$ and $K$ should 
approach zero with increasing distance $b$ to the wind 
source. However, our calculations have shown that they do so only very 
slowly, decreasing as the inverse of the logarithm of the distance to 
the source. This opens the question of whether, in fact, when the flow 
reaches the terminal shock separating it from the outer medium, $I$ has indeed 
vanished. 
 
If not, the region in the wind cavity at $r \gg r_A$ would be in an 
intermediate asymptotic regime, i.e., one in which $I$ is still 
finite, though smaller than $I_{sup}$. For classical winds, the flux 
distribution would then be as given by Eq.(\ref{soltgpsiofa}), with the value of 
$\rho_0(b)$ and $I(b)$ in the asymptotic domain given by 
Eqs.(\ref{bennetclass}) and (\ref{matchingrelation}).  Extreme 
focusing would be possible for $I$ close to its maximum value, $I_{sup}$, 
defined in Eq.(\ref{alphaEoverOmegaclass}). However such a wind
would not yet have made its transition through the fast 
critical point. Similar considerations 
apply to relativistic winds. 
 
Such an intermediate asymptotic state is associated with a finite value 
of $I < I_{sup}$ (or $K < K_{sup}$).  It is appropriate to discuss the 
properties of winds which carry a current close to the maximum 
possible value, since these have strong focusing properties and may be 
directly related to ubiquitous jet phenomena.  We show below that such 
winds have very special properties.  For example, when the current is 
close to $I_{sup}$, the wind splits into a focused jet and an 
equatorial wind with radial geometry. In between these two components, 
there is only a very weak flow, but a large Poynting flux. Such highly 
focused winds must be sub fast mode as we show below.

\subsection{Poynting-Flux Dominated Winds: Polar Jets and Equatorial Winds} 
\label{vinfinityzero} 
 
We now discuss the structure of rotating MHD winds which carry a 
circumpolar current close to the maximum possible value. This current 
is treated as remaining constant in the intermediate region as 
discussed above. The wind's structure is given by 
Eq.(\ref{soltgpsiofa}).  Suppose that $I$ is close to its maximum 
possible value, $I_{sup}$, the minimum of the function $(\alpha E 
/\Omega)$, reached at some $a_* \neq 0$.  It is assumed that $(\alpha 
E /\Omega)$ has such a minimum.  In this case, the integral in 
Eq.(\ref{soltgpsiofa}) would almost diverge at $a_*$.  As a result, 
all the flux at $ a \le a_* $ is strongly focused about the axis. The 
focusing is still conical however because divergence is not exactly 
reached. The flux at $a \ge a_*$ is spread about the equator. The 
terminal flow velocity in the vicinity of $a = a_*$ almost vanishes, 
while the angular extent of this region is large. 
This structure is represented in Fig.(\ref{fig2}). 
 
A wind close to the maximum current then appears to have a rather 
peculiar structure. It is split into two main flow regions, a 
circumpolar flow channeled in a cone of very small opening angle, and 
a conically spread equatorial flow.  These two regions are separated 
by a region with very little flow, where the outflowing energy is 
almost entirely in Poynting form. This result has been also obtained 
in numerical simulations by \citet{Ustyugovaetal00} of the flow from a 
Keplerian disk threaded by a dipole-like magnetic field. 
 
Again, from the discussion of section (\ref{sectionfill}), this regime 
cannot be the eventual asymptotic regime. The latter necessarily 
involves a vanishing circumpolar current. The solutions described 
above occur in the intermediate asymptotic regime , which may persist 
over large distances. That $I$ or $K$ are close to their maximum 
possible value, $I_{sup}$ or $K_{sup}$, has however here the character 
of an extra and rather conjectural assumption. 
 
In the particular case when the function $(\alpha E/\Omega)$ reaches 
its minimum at $a = 0$, the situation regarding the focusing of the 
flow is different. The flux function $a$ is proportional to $r^2$ near 
the polar axis. Symmetry about the axis implies that any physical 
quantity be even in $r$, and reaches a flat extremum at $r=0$. This 
however does not imply that functions of the flux $a$ should behave 
similarly. On the contrary, a linear variation with $a$ should be the 
rule. Therefore when $a_* = 0$ and the circumpolar current is 
supposedly close to $\alpha_0 E_0/\Omega_0$, there is no divergence of 
$\tan{\psi(a)}$ (Eq.(\ref{soltgpsiofa})), as $a$ approaches 
zero, due to the vanishing of the square root in the denominator, 
because the latter usually has a simple zero.  Most of the mass flux 
is then spread about the equator in a wind shaped like a conical fan. 
This structure is represented in Fig.(\ref{fig3}). 
There is little flow about the axis, since the velocity, equal to 
$\sqrt{2(E - I \Omega /\alpha)}$, is small in this region. 
Eq.(\ref{soltgpsiofa}) shows that the wind nevertheless fills all 
space. Indeed, $\psi(a)$ approaches $\pi/2$ as $a$ approaches zero 
because $I$ vanishes on the axis.  The flow near the axis would in 
this case be weak, the energy being transported mainly in 
electromagnetic form in a circumpolar flux tube subtending negligible 
flux. 
 
\subsection{ Neutral Magnetic Surfaces do not Focus Winds} 
\label{secborderedbyneutral} 
 
In section (\ref{vinfinityzero}) above, a wind carrying nearly maximal 
current in an intermediate asymptotic regime has been considered. 
This situation induces a strong focusing of magnetic surfaces. Note 
that at a neutral surface, where $I$ also vanishes, 
Eq.(\ref{soltgpsiofa}) does not give rise to a divergence of $\tan 
\psi(a)$, since the integral in Eq.(\ref{soltgpsiofa}) is convergent 
\citep{HN2001a,HN2001b}.  Neutral surfaces do not force the 
magnetic surfaces nested in them to cylindrical shapes.  It is shown 
in Appendix(\ref{appdeconfinecylindrical}) that a neutral magnetic 
surface at the jet's edge does not cause the radius of a 
pressure-confined jet to diverge.  We conclude that currents returning 
at neutral magnetic surfaces do not cause any general focusing of the 
structure. \citet{Okamoto99} shows that they rather have a defocusing effect. 
 
\section{ Conclusions on the Asymptotic Regime of MHD Winds} 
\label{sectionconclusion}

The major conclusions of this paper are given below. 
 
\begin{enumerate} 
\item{ If there were to be a finite circumpolar current at infinity then, for
 classical winds, this current should have a value equal to the
 minimum value of the function $ \alpha E / \Omega $.  This minimum
 value, $I_{sup}$, is the maximum allowed current.  If this maximum
 current is achieved,then there exists a particular magnetic surface,
 $a_*$, where $ \alpha E / \Omega $ reaches its minimum and where the
 total energy is carried by Poynting flux alone.}
 
\item{ The limiting maximum current, $I_{sup}$, can, however, only be reached 
      from below. In other words, as the distance from the source
      increases, the circumpolar current would have to increase. This
      cannot be achieved fro winds emanating from a finite, central
      source of magnetic flux.  Thus, we conclude that, for
      non-pathological cases, the maximal current cannot be reached
      from below. Therefore, the asymptotic circumpolar current must
      vanish.}
	 
\item{ A winds carrying a cicumpolarcurrent, $I$, that is bounded from below
 such that, $I > {2 \over 3} I_{sup}$, cannot have a terminal velocity
 everywhere in excess of the fast mode speed.}
 
\item{ The circumpolar current declines to zero only inversely 
	logarithmically with distance.  An important consequence of
	this very slow decline is that there exists an extensive
	intermediate asymptotic regime, where this current is still
	finite.  Thus, significant Poynting flux can be carried over
	very large distances. The termination of the wind, for example
	at a terminal shock, may occur before this current has decayed
	to zero and the $\sigma$ parameter is still large}
\end{enumerate} 
 
Similar conclusions apply to relativistic winds: 
 
\begin{enumerate} 
 
\item{ The total circum-polar proper current $K$ is, for relativistic 
	winds, less than $K_{sup}$, the minimum value of the function
	$ \alpha (E-c^2)/ \Omega $. Any finite value of this proper
	current at infinity equals $K_{sup}$.  If this maximum current
	were to be achieved asymptotically, there would exist a
	particular magnetic surface, $a_*$, where $ \alpha(E - c^2)/
	\Omega $ reaches its minimum value and where the total energy
	is carried by Poynting flux alone.  }
 
\item{ The upper bound to the proper current, $K_{sup}$, can only be reached
 from below. As in the non-relativistic case, it is not physically
 possible to reach this upper bound for winds emanating from a finite,
 central source of magnetic flux.  Thus, the circum-polar proper
 current should formally approach zero at infinity.}
 
\item{ Winds that carry a proper current such that inequality 
	(\ref{ineqforsuperfastrelat}) is not satisfied cannot have 
	terminal velocities everywhere in excess of the fast mode 
	speed.  } 
 
\item{ The circumpolar total proper current declines to zero only 
       inversely logarithmically with distance.  A consequence of this 
       very slow decline is that there is an extensive intermediate 
       asymptotic regime where this current is still finite. 
       Significant Poynting flux can be carried over large distances, 
       and the wind may terminate in a shock much before this current 
       has decayed to zero.} 
 
\item{ In the intermediate asymptotic regime, a classical or 
	relativistic wind carrying a proper current close to the 
	maximum allowed value $I_{sup}$ (or $K_{sup}$) consists of a 
	conical wind of very small opening angle about the pole 
	surrounded by a region in which most of the energy flux is in 
	Poynting form.} 
 
\item { Such close-to-maximum Poynting Flux carrying winds with polar 
	jets and equatorial winds must be sub-fast mode. External 
	boundary conditions may significantly influence their 
	nature. }

\item{ For force-free initial conditions in the near field, the energy 
	close to the source is all Poynting flux. Thus, relativistic 
	jets emanating from a strongly magnetized base can maximize 
	the output in Poynting flux} 
 
\end{enumerate} 
 
\acknowledgements 
 
The authors thank the Space Telescope Science Institute and the Johns
Hopkins University for continued support to their collaboration over
the last decade. J.H also thanks the EC Platon program
(HPRN-CT-2000-00153) and the Platon collaboration. CN is pleased to
thank the Director of ESO for support and hospitality during which
time this paper was completed.  We thank Sundar Srinivasan for
significant help with the figures.
 
\bigskip 
\bigskip 
\newpage 
  
\appendix 
 
\renewcommand{\theequation}{A-\arabic{equation}} 
\setcounter{equation}{0} 
\section{Unconfined Jets as limits of cylindrical pressure-confined jets} 
\label{appdeconfinecylindrical} 
 
We analyze the structure of a pressure-confined cylindrical jet 
with a given set of first-integrals as the confining pressure is 
reduced.  For completeness, we consider both the case of a 
jet bordered by an ordinary 
magnetic surface and the case of a jet enclosed by 
a neutral magnetic surface. This is appropriate  
since the electric current organizes 
itself into closed cells and because the presence of a neutral 
magnetic surface at the jet's edge slightly changes its structure, 
favoring space filling. 
 
Where a uniform external pressure imposes cylindrical geometry for the 
magnetic surfaces, the orthogonal trajectories are straight lines 
perpendicular to the rotation axis. Taking into account the 
smooth matching with the polar boundary layer the radius $r(a)$ of the 
cylindrical flux surface $a$ is found to be: 
\begin{equation} 
r^2(a) = 
{{\Gamma }\over{\Gamma -1}} \ 
{{ Q_0 \mu_0 \alpha_0^2 \rho_0^{\Gamma - 2} }\over{ 
\Omega_0^2}} \exp \left( \int_0^a 
{{ \sqrt{2} \Omega(a') da'}\over{ 
\mu_0 I \sqrt{E(a') - I \Omega(a') /\alpha(a') } 
}} \right) 
\label{rversusaclass} 
\end{equation} 
The variables $I$ and $\rho_0$ are  related by the Bennet relation  
\begin{equation} 
{{\Gamma }\over{\Gamma -1}} \ 
Q_0 \rho_0^{\Gamma - 1} = {{I \Omega_0}\over{ \alpha_0}}  
\label{bennetcylindrical} 
\end{equation} 
They both depend on the external confining pressure, 
$P_{ext}$, as does also the radius $r_*$ of the outermost magnetic 
surface. The boundary condition imposes equilibrium between the 
externally applied pressure and the total pressure at the edge of the 
jet. The latter is the sum of toroidal magnetic pressure and gas pressure 
since poloidal magnetic pressure can be  neglected.  If $a_*$ is simply an 
ordinary magnetic surface, gas pressure can also be neglected and the 
boundary condition becomes: 
\begin{equation} 
P_{ext} = {{\mu_0 I^2}\over{ 2 \ r_{*}^2}} 
\label{BCconfinement} 
\end{equation} 
where $r_*$ depends on  $I$ as given by Eq.({\ref{rversusaclass}). 
In this case, $I$ and $P_{ext}$ are simply related by  
\begin{equation} 
2 \left({{\Gamma}\over{\Gamma -1}}\right)^{ {{1}\over{\Gamma -1}} } 
\left({{\alpha_0 Q_0 \rho_{A0}^{\Gamma -1} 
}\over{ I \Omega_0}}\right)^{ {{\Gamma}\over{\Gamma -1}} } 
\exp \left( \int_0^{a*} 
{{\Omega(a') da' 
}\over{ 
\sqrt{2} \mu_0 I \sqrt{E(a') - I \Omega(a') /\alpha(a') } 
}} \right) = {{Q_0 \rho_{A0}^{\Gamma} }\over{ P_{ext}}} 
\label{eqIdeconfine} 
\end{equation} 
As $P_{ext}$ decreases, the left hand side of 
Eq.(\ref{eqIdeconfine}) should diverge, which leaves at most two solutions
for $I$, $I =0$ and, if $(\alpha E/\Omega)$ has a flat minimum, $I= I_{sup}$ 
(\citet{HN2001a}, \citet{HN2001b}, \citet{LeryetalII}).  The value 
$I_{sup}$ is reached by increasing $I$ as $P_{ext}$ decreases. 
 
Some complications arise when the surface $a_*$ is a neutral one. The jet 
is then bordered by a neutral surface.  The proof that 
$I =0$ or $I_{sup}$ in this  unconfined limit 
should be extended to this case. 
The outskirts of the pressure-confined jet then consist of half a 
sheet pinch boundary layer, in which the transfield equilibrium 
reduces approximately, $\vec{n}$ being the unit vector 
normal to magnetic surfaces, to: 
\begin{equation} 
\vec{n} \cdot \vec{\nabla} \ 
\left( Q \rho^{\Gamma} + {{B_{\theta}^2}\over{2 \mu_0}} \right) = 0 
\label{transnullpinch} 
\end{equation} 
 $B_{\theta}$ is given by Eq.(\ref{rBthetaass}). Integrating accross 
the boundary layer, we get, taking into account the divergence of the 
function $\alpha$ at the null surface: 
\begin{equation} 
Q \rho^{\Gamma} + {1 \over 2} 
{{\rho^2 r^2 \Omega}\over{ \mu_0 \alpha^2}} 
= Q_* \rho_*^{\Gamma} 
\label{transnullinteg} 
\end{equation} 
where $\rho_*$ is the density at the jet's edge. It does not vanish, 
because of the confining pressure.  Treating $Q$, $E$ and $\Omega$ as 
constants equal to $Q_*$, $E_*$ and $\Omega_*$ in this external 
boundary layer, Eq.(\ref{transnullinteg}) provides an expression for 
$\alpha$ as a function of the parameter 
\begin{equation} 
X = {{\rho}\over{\rho_*(b)}} 
\label{paraX*} 
\end{equation} 
namely: 
\begin{equation} 
{{1}\over{\mu_0 \alpha^2}} = 
{{2 Q_* \rho_*^{\Gamma -2} 
}\over{ 
\Omega_*^2 r^2   }} 
\left( {{1}\over{X^2}} - {{1}\over{X^{2 - \Gamma} }} \right) 
\label{alphaofX*} 
\end{equation} 
Eq.(\ref{alphaofX*}) implicitly gives $a$ as a function of $X$  
for known $\alpha(a)$. The function $\alpha$ 
 cannot be treated as approximately constant because of its divergence 
 at $a = a_*$. Nevertheless, in our analysis, $\alpha$ is a given 
 function of $(a_* - a)$ in this vicinity.  It is expected that $r$ 
 itself should vary only little in the boundary layer if the latter is 
 indeed thin, which will be checked {\it a posteriori}.  The flux 
 distribution versus radius results from Eq.(\ref{defalpha}), the 
 velocity being obtained from Eq.(\ref{Bern}). Disregarding the 
 gravitational potential and the toroidal velocity and assuming that 
 the velocity at the jet's edge is largely supersonic, we obtain 
\begin{equation} 
r {{dr}\over{da}} = {{\alpha (a) }\over{ \rho_* \ X \ \sqrt{2 E_*} }} 
\label{fluxdistin*BL} 
\end{equation} 
Eq.(\ref{fluxdistin*BL}) implicitly gives $r(X)$ 
and Eq.(\ref{alphaofX*}) implicitly gives $a(X)$.  To derive an 
explicit parametric representation of the flux radius relation in the 
boundary layer we deduce $a(X)$ from Eq.(\ref{alphaofX*}).   
We have shown \citep{HN2001a} that in the vicinity of 
the neutral surface $\alpha(a)$ varies as: 
\begin{equation} 
{{1}\over{ \alpha(a)}} = {{ (a_* - a)^{\nu} }\over{ \eta a_*^{\nu} }} 
\label{alphanearneutral} 
\end{equation} 
where $\eta$ is a constant having the same dimension as $\alpha$ and 
$\nu$ is a positive exponent, strictly smaller than unity and most 
often equal to $(1/2)$.  This implies that the edge boundary layer is not 
infinitely extended. Indeed, as $a$ approaches $a_*$, $X$ becomes 
close to unity and, in this limit, the relation between radius and 
$\alpha$ as given by Eqs.(\ref{fluxdistin*BL}) and 
(\ref{alphanearneutral}) reduces to: 
\begin{equation} 
r \ dr \sim \alpha^{-1/\nu} \ d\alpha 
\label{rversusalphaapprox} 
\end{equation} 
Since $(1/\nu)$ is strictly larger than unity, this implies that, as 
$\alpha$ diverges near $a_*$, $r$ remains bounded. This result is 
similar to our finding \citep{HN2001a} that the angular thickness of a 
neutral sheet boundary layer remains limited and in fact small. 
Assume, similarly, that the boundary layer thickness remains small in 
this case, so that $r$ varies only little.  The boundary layer 
thickness is then obtained, using Eqs.(\ref{alphaofX*}) and 
(\ref{alphanearneutral}), by turning Eq.(\ref{fluxdistin*BL}) into the 
following differential equation for $r(X)$: 
\begin{equation}  
2 r {{dr}\over{dX}} = {{\eta a_*} \over{ \rho_* \sqrt{2 E_*} }} 
\left( {{ 2 Q_* \mu_0 \eta^2 \rho_*^{\Gamma - 2} }\over{  
\Omega_* r_*^2 }} \right)^{ {1 \over 2 \nu} - {1 \over 2} }  
\ \left( {{2 - (2 - \Gamma) X^{\Gamma} }\over{ 
X^{ {1 \over \nu}} (1 - X^{\Gamma} )^{{3\over 2} - {1 \over 2 \nu} } }} \right) 
\label{diffequationforrofX} 
\end{equation} 
Integrating, the thickness $\Delta_*$ of the edge boundary layer is 
found. Let us denote by  $k(\nu, \Gamma)$ the integral 
\begin{equation} 
k(\nu, \Gamma) = \int_{{1\over 2}}^{1}  
{{ \left(  2 - (2 - \Gamma) X^{\Gamma}\right) \ dX }\over{ 
X^{ {1 \over \nu}} (1 - X^{\Gamma} )^{{3\over 2} - {1 \over 2 \nu} } }} 
\label{definekofnuandgamma} 
\end{equation} 
which is convergent at $X = 1$ since 
$\nu$ is less than unity. From Eq.(\ref{diffequationforrofX}) we get: 
\begin{equation} 
2 r_* \Delta_* = {{ k(\nu, \Gamma) \eta a_*} \over{ \rho_* \sqrt{2 E_*} }} 
\left( {{ 2 Q_* \mu_0 \eta^2 \rho_*^{\Gamma - 2} }\over{ 
\Omega_* r_*^2 }} \right)^{{1 \over 2 \nu} - {1 \over 2} } 
\label{edgeBLthickness} 
\end{equation} 
The total outer radius $r_*$ of the jet, when under the pressure $P_{ext}$, 
is obtained by adding $\Delta_*$ to the extent of the domain occupied 
by the flux tube at the inner limit of the edge boundary 
layer.  Since this flux is close to being $a_*$, this radius is almost 
given by Eq.(\ref{rversusaclass}) with $a = a_*$. The total poloidal 
current is treated as if it were constant up to $a_*$, whereas in 
fact it decreases to zero in the boundary layer.  Then we find: 
\begin{eqnarray} 
r_*^2 =  
{{\Gamma }\over{\Gamma -1}} \ 
{{ Q_0 \mu_0 \alpha_0^2 \rho_0^{\Gamma - 2} }\over{ 
\Omega_0^2}}  \exp \left( \int_0^{a_*} 
{{ \sqrt{2} \Omega(a') da'}\over{ 
\mu_0 I \sqrt{E(a') - I \Omega(a') /\alpha(a') } 
}} \right) \nonumber \\  
+  {{ k(\nu, \Gamma) \eta a_*} \over{ \rho_* \sqrt{2 E_*} }} 
\left( {{ 2 Q_* \mu_0 \eta^2 \rho_*^{\Gamma - 2} }\over{ 
\Omega_* r_*^2 }} \right)^{{1 \over 2 \nu} - {1 \over 2} } 
\label{outerradiuswithedgeBL} 
\end{eqnarray} 
The edge density $\rho_*$ is related to $P_{ext}$ by  
the boundary condition 
\begin{equation} 
P_{ext} = Q_* \rho_*^{\Gamma} 
\label{BCforpressurizedjet} \end{equation} 
The neutral layer radius is related to $P_{ext}$ and to $I$ 
by the neutral surface's Bennet relation, 
\begin{equation} 
\mu_0 I^2 = 2 Q_* \rho_*^{\Gamma} r_*^2 
\label{bennetatedge} 
\end{equation} 
The axial density $\rho_0$  is related to  
$I$ by Eq.(\ref{bennetcylindrical}). 
Eliminating $\rho_*$, $\rho_0$ and $r_*$ in favour of $I$, 
Eq.(\ref{outerradiuswithedgeBL}) takes the form of 
a relation between $I$ and $P_{ext}$, which is best written  
by introducing the dimensionless variables $D$ and $i$ defined by: 
\begin{equation} 
\rho_* = \mu_0 \alpha_0^2  D 
\label{defD} 
\end{equation} 
\begin{equation} 
I = {{\alpha_0 E_0}\over{\Omega_0}} i 
\label{deflittlei} 
\end{equation} 
The relation between $I$ and $P_{ext}$ then translates 
into the following relation between $i$ and $D$: 
\begin{eqnarray} 
{{2Q_*}\over{Q_0}}  
\left({{\Gamma}\over{\Gamma -1}}\right)^{{{1}\over{\Gamma -1}}} 
\left( {{Q_0 \rho_{A0}^{\Gamma -1} }\over{i \ E_0}}  
\right)^{{{\Gamma}\over{\Gamma -1}}}  
\exp \left( {{ \sqrt{2} \Omega_0^2 a_* }\over{  
\mu_0 \alpha_0 E_0^{3/2} }} 
\int_0^{a_*}  
{{ \Omega(a') da'}\over{ i \  \Omega_0 a_* 
 \sqrt{ {{E(a')}\over{E_0}}  
- i \ {{\Omega (a') \alpha_0}\over{\Omega_0 \alpha(a')}} }    
}} 
\right)  
\nonumber \\ 
+ {{    2^{ {{2 - \nu}\over{\nu}} }   
k(\nu, \Gamma)  a_* \Omega_0^2 }\over{ 
\mu_0  \alpha_0^2 \ E_0^{3/2}   }} 
\left( {{E_0}\over{E_*}}\right)^{1 \over 2} 
\left( {{\eta}\over{\alpha_0}} \right)^{1 \over \nu}   
\ \left({{\Omega_0}\over{ \Omega_*}} \right)^{ {1 \over \nu} - 1}  
\ \left({{Q_* \rho_{A0}^{\Gamma -1} }\over{  E_0}} \right)^{1 \over \nu}   
\ {{ D^{ {{\Gamma -1}\over{\nu}} - \Gamma }   }\over{ 
i^{ {1 \over \nu} + 1}      }}  = {{1}\over{D^{\Gamma}}} 
\label{bigequationforiofD} 
\end{eqnarray} 
which is of the form: 
\begin{equation} 
\Lambda_0 \ {{\Phi(i) }\over{i^{ \Gamma/(\Gamma -1)}  }}  
+ \Lambda_1 {{ D^{ {{\Gamma -1}\over{\nu}} - \Gamma } }\over{  
i^{ {1 \over \nu} + 1}      }} = {{1}\over{D^{\Gamma} }} 
\label{formofbigequation} 
\end{equation} 
The coefficients $\Lambda_0$ and $\Lambda_1$ are positive and $\Phi(i)$ is: 
\begin{equation} 
\Phi(i)  = \exp \left( {{ \sqrt{2} \Omega_0^2 a_* }\over{ 
\mu_0 \alpha_0 E_0^{3/2} }} 
\int_0^{a_*} 
{{ \Omega(a') da'}\over{ i \  \Omega_0 a_* 
 \sqrt{ {{E(a')}\over{E_0}} 
- i \ {{\Omega (a') \alpha_0}\over{\Omega_0 \alpha(a')}} } 
}} 
\right) 
\label{defPhiofi} 
\end{equation} 
When the external pressure becomes very small, the second term on the 
left-hand side of Eq.(\ref{formofbigequation}) becomes negligible if 
$i$ remains finite and a non-vanishing solution is then found for $i = 
i_{sup}$, the value of $i$ that causes the function $\Phi(i)$ to 
diverge. If $\Phi(i)$ does not diverge for finite $i$, the solution for 
$i$ as $D$ becomes small moves towards $i = 0$. This solution exists 
also when $\Phi(i)$ diverges for finite $i$.  The second term on the 
left hand side of Eq.(\ref{formofbigequation}) 
is then again negligible with respect to the first one because 
of the very rapid divergence of $\Phi(i)$ as $ i$ approaches zero. The 
solution is then close to that of the equation 
\begin{equation} 
{{1}\over{D^{\Gamma} }} =  \Lambda_0 
{{\Phi (i) }\over{i^{ \Gamma/(\Gamma -1)}  }}  
\label{zeroilimitofbigeq} 
\end{equation} 
We conclude that in this limit of vanishing external pressure 
the outer boundary layer has negligible   
effect on the relation between the current  
and the external pressure. Actually, Eq.(\ref{zeroilimitofbigeq}) 
is just the non-dimensional form of Eq.(\ref{rversusaclass}). 
As a result $I =0$ and $I = I_{sup}$  
remain the only possible solutions in the limit of  
deconfinement, even when the jet is bordered by a  
neutral magnetic surface. 
 
\bigskip 
\newpage

\renewcommand{\theequation}{B-\arabic{equation}} 
\setcounter{equation}{0} 
\section{Fast Critical Point Position} 
\label{appendcritatinftyclass} 
  
For conical asymptotics the function $ \vert r \nabla a \vert $ is 
bounded from above and below with $ \mu(a) \leq \vert r \nabla a \vert 
\leq \lambda(a)$ as $(r / r_A)$ tends to infinity \citep{HN89}. Thus, 
for conical asymptotics $(d (r \vert \nabla a \vert) / d s)$ 
approaches $0$.  We assume that there is only one fast  
point in this analysis.  For 
cylindrical asymptotics, $(d \vert r \nabla a \vert / d s)$ 
also approaches $0$ and the fast point may be at infinity.   
For parabolic field lines, 
representable as by $z = K(a) r^{p(a)}$, say,  
$ r \vert \nabla a \vert$ is given by: 
\begin{equation} 
 r \vert \nabla a \vert = {{\sqrt{p^2(a) + r^2/z^2}}\over{\vert{K'/K} 
 + p' log r\vert}} 
\end{equation} 
For non-constant $p(a)$, $r \vert \nabla a \vert$ approaches  
$0$, as $r$ approaches 
infinity. If $p(a)$ is a constant larger 
than unity, $r \vert \nabla a \vert$ approaches a constant value   
and $ d(r \vert \nabla a \vert)/ds$ approaches zero. If $p(a) = 1$, $r/z$ 
goes to a constant as $s$ approaches infinity, and again  
$ d(r \vert \nabla a \vert)/ds$ approaches $0$.  
In general,  $ d(r \vert \nabla a \vert)/ds$ becomes very small 
at large distances, allowing a fast critical point  
to be located in the asymptotic region.

\clearpage 
 
\begin{figure} 
\plotone{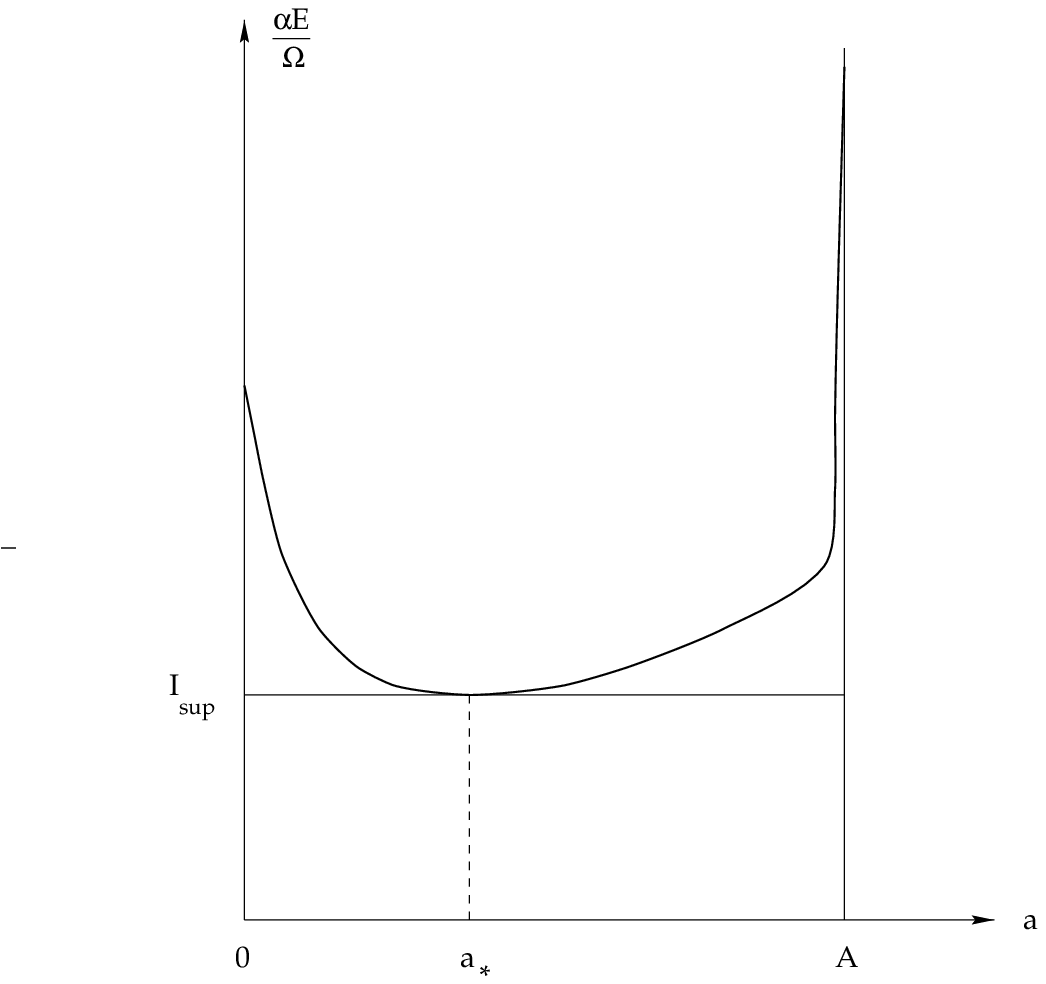} 
\caption{ The function $(\alpha E/\Omega)$ versus flux $a$ 
for some classical wind and for a poloidal field of 
supposedly dipolar symmetry. 
The equatorial plane, at $a = A$, 
is the only null surface. 
Both $\alpha$ and $(\alpha E/\Omega)$ 
diverge at a null surface.  
The current $I$ (Eq.(\ref{defI})) 
can only take values between 
zero and $I_{sup}$, the minimum value of 
$(\alpha E/\Omega)$. 
This function is assumed here to 
have an absolute minimum at some $a_*$  
away from the polar axis. 
The value of $I$ at infinitely large 
distances could then be zero or $I_{sup}$  
(see however sections (\ref{subsecalphaEoverOmegaclass}) and  
(\ref{subsecalphaEoverOmegarelat})). 
The flux contained in a cylindrical jet about the axis at 
such large distances  must be $2 \pi a_*$  
(section \ref{consistencyconescylinders}). 
For some winds, the function  $\alpha E/\Omega$ could 
be monotonically increasing. In this case 
$I$ could only vanish at infinitely large distances. 
For relativistic winds, the function $(\alpha (E - c^2)/\Omega)$ 
should be substituted for $(\alpha E/\Omega)$ 
and the proper current $K$ (Eq.(\ref{IandK})) for $I$, with no 
other change in the above analysis 
\label{fig1}} 
\end{figure} 
 
\clearpage 
 
\begin{figure} 
\epsscale{1}\plotone{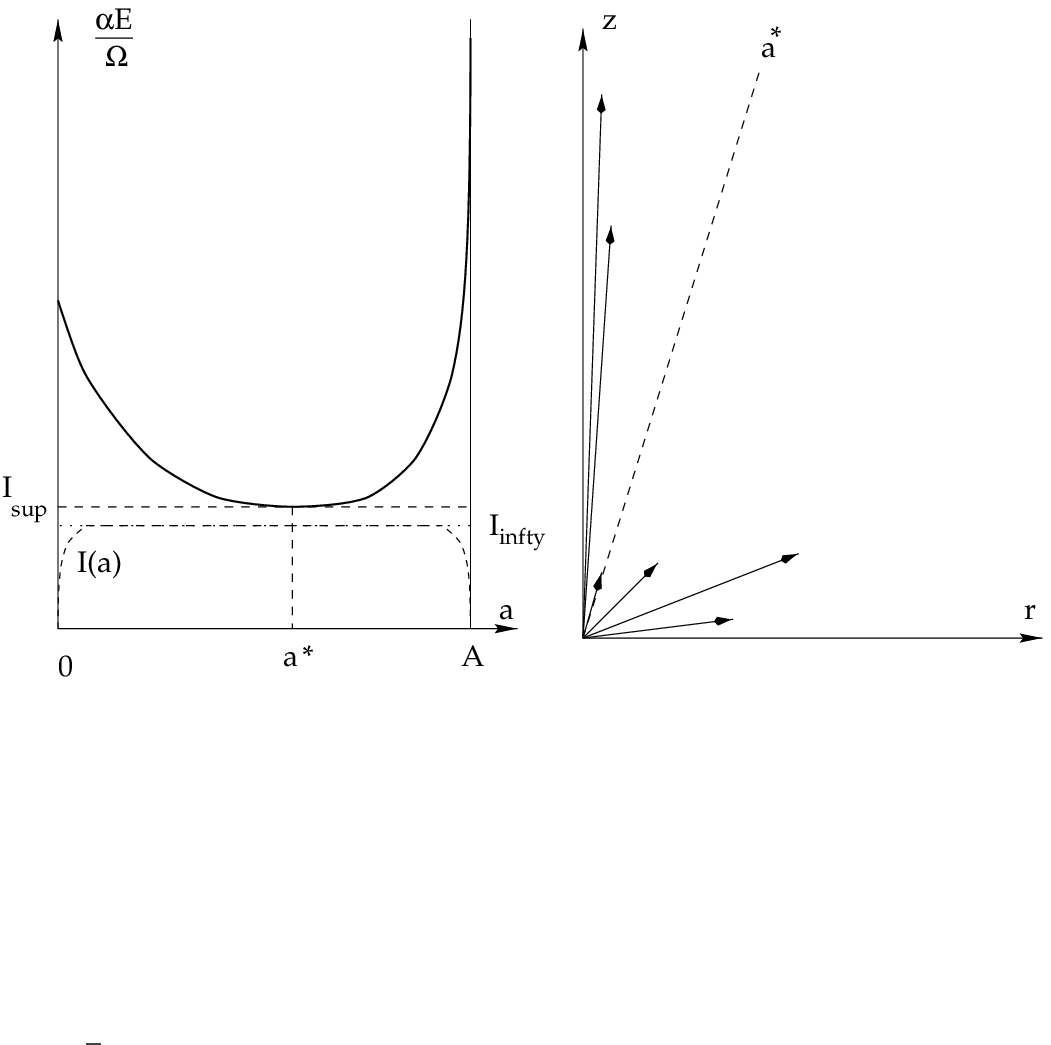} 
\caption{A near-maximal current jet. In an intermediate 
asymptotic regime, the circum-polar current, $I_{\infty}$, 
may still be close to the maximal value $I_{sup}$. The left panel 
represents $(\alpha E/\Omega)$  
(or $(\alpha (E - c^2)/\Omega)$  for a relativistic wind), 
$I_{sup}$ and $I_{\infty}$  
(or $K_{sup}$ and $K_{\infty}$ resp.), as well as the run with $a$ 
of the enclosed total current $I(a)$ (resp. $K(a)$),  
taking into account the polar ($a =0$) and equatorial 
($a=A$) current-carrying boundary layers.  
The wind structure (Eq.(\ref{soltgpsiofawithoutneutrallayers})  
is represented  
in the right panel. Each arrow indicates the wind  
speed and direction for a set 
of equidistant values of the flux  
variable $a$. The dashed line represents the  
magnetic surface $a = a_{*}$. The regions close to the pole and the equator 
are boundary layers where the circum-polar 
current and return current flow. These bounday layers  
are not explicitly represented.  
\label{fig2} } 
\end{figure} 
 
\clearpage 
 
\begin{figure} 
\epsscale{1}\plotone{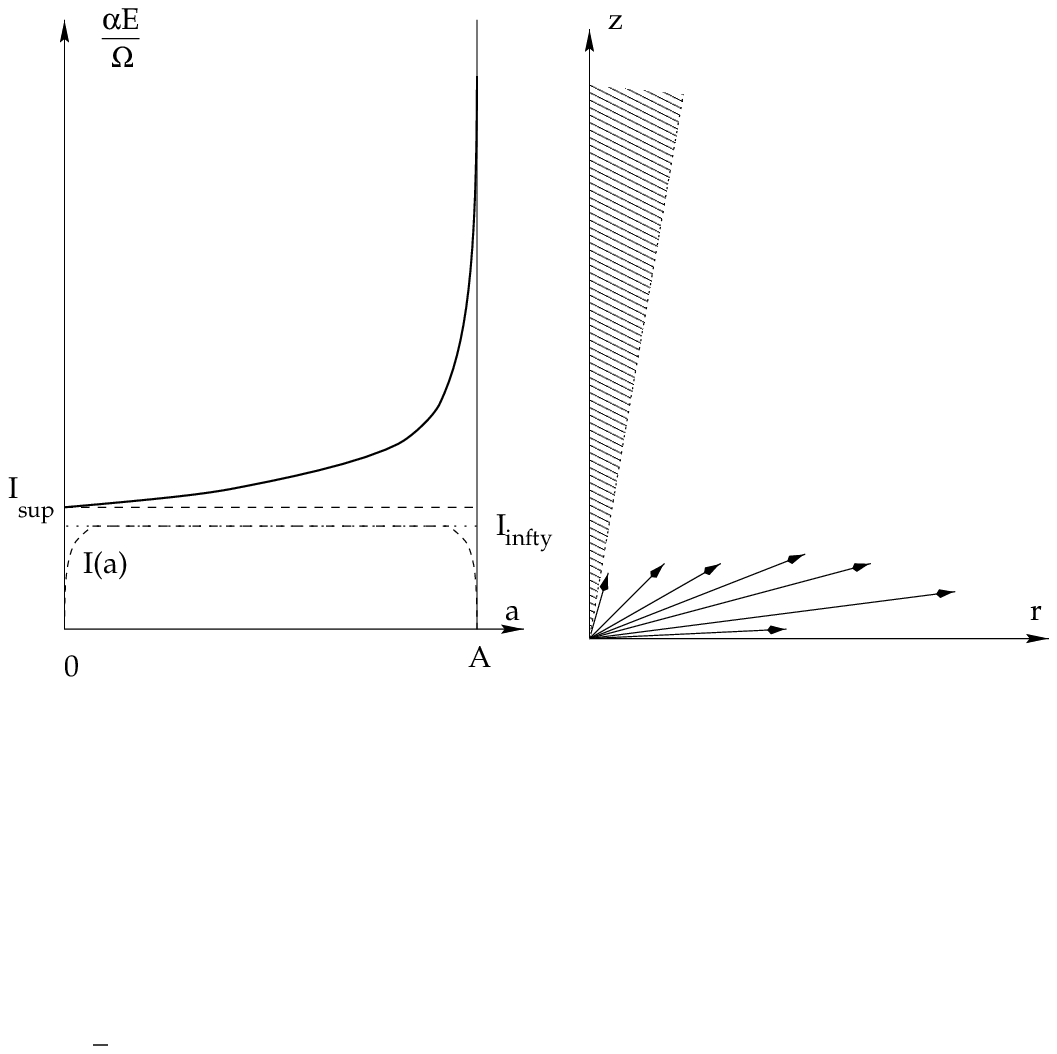} 
\caption{ A near-maximal current wind. In an intermediate 
asymptotic regime,the circum-polar current, $I_{\infty}$, 
may still be close to the maximal value $I_{sup}$. 
In the present case, this maximum value is supposedly  
reached at the polar axis, $a =0$. The left panel 
represents $(\alpha E/\Omega)$  
(or $(\alpha (E - c^2)/\Omega)$  for a relativistic wind), 
$I_{sup}$ and $I_{\infty}$ 
(or $K_{sup}$ and $K_{\infty}$ resp.), as well as the run with $a$ 
of the enclosed total current  $I(a)$ (resp. $K(a)$), 
taking into account the polar ($a =0$) and equatorial 
($a=A$) current-carrying boundary layers.  
The wind structure (Eq.(\ref{soltgpsiofawithoutneutrallayers}))  
is represented  
in the right panel. Each arrow indicates the wind  
speed and direction for a set 
of equidistant values of the flux  
variable $a$. The hatched region is the  
boundary layer about the polar axis.  
\label{fig3} } 
\end{figure} 
 
\clearpage

\end{document}